\RequirePackage{ifpdf}
\ifpdf % We are running pdfTeX in pdf mode
\documentclass[pdftex]{sigma}
\else
\documentclass{sigma}
\fi

\begin{document}
\allowdisplaybreaks

\renewcommand{\PaperNumber}{027}

\FirstPageHeading

\ShortArticleName{Anomalously Slow Cross Symmetry Phase
Relaxation}

\ArticleName{Anomalously Slow Cross Symmetry Phase Relaxation,
Thermalized Non-Equilibrated Matter and Quantum Computing Beyond
the Quantum Chaos Border}

\Author{M. BIENERT~$^\dag$, J. FLORES~$^\dag$, S.Yu.
KUN~$^{\dag,\ddag,\S}$ and T.H. SELIGMAN~$^\dag$}

\AuthorNameForHeading{M. Bienert, J. Flores, S.Yu. Kun and T.H.
Seligman}

\Address{$^\dag$ Centro de Ciencias F{\'i}sicas, Universidad
Nacional
  Aut{\'o}noma de M{\'e}xico,\\
  $\phantom{^\dag}$~Cuernavaca, Morelos, Mexico}
\EmailD{\href{mailto:bienert@fis.unam.mx}{bienert@fis.unam.mx},$\!$
\href{mailto:jfv@servidor.unam.mx}{jfv@servidor.unam.mx},$\!$
\href{mailto:kun@fis.unam.mx}{kun@fis.unam.mx},$\!$
\href{mailto:seligman@fis.unam.mx}{seligman@fis.unam.mx}$\!$}

\Address{$^\ddag$ Nonlinear Physics Centre, RSPhysSE, ANU,
Canberra ACT 0200, Australia}

\Address{$^\S$ Department of Theoretical Physics, RSPhysSE, ANU,
Canberra ACT 0200, Australia}

\ArticleDates{Received November 30, 2005, in f\/inal form February
08, 2006; Published online February 27, 2006}

\Abstract{Thermalization in highly excited quantum many-body
system does not necessa\-rily mean a complete memory loss of the way
the system was formed. This ef\/fect may pave a~way for a quantum
computing, with a large number of qubits $n\simeq 100$--1000, far
beyond the quantum chaos border. One of the manifestations of such
a thermalized non-equilibrated matter is revealed by a strong
asymmetry around 90$^\circ $ c.m. of evaporating proton yield in
the Bi($\gamma$,p) photonuclear reaction. The ef\/fect is
described in terms of anomalously slow cross symmetry phase
relaxation in highly excited quantum many-body systems with
exponentially large Hilbert space dimensions. In the above
reaction this phase relaxation is
 about eight orders of magnitude slower than energy relaxation (thermalization).}


\Classification{82C10; 81V35; 81Q50; 34L25}

\section{Introduction}

Independent-particle modes in interacting many-body systems result
from a mean-f\/ield appro\-ximation and are at the center of many
theoretical considerations. In the case of a quantum information
device this independent-particle basis can be considered as the
basis spanned by the individual qubits, the ``computational
basis''. At high excitation energy, the interaction between the
particles results in a rapid mixing of the independent particle
states \cite{wigner55:57,wigner72,anderson,GM-GW98}. This mixing
leads to the formation of complicated many-body conf\/igurations.
Each of these individually ergodic (independent of the initial
conditions) many-body states is characterized by sharing the
energy between many particles of the system. The characteristic
time for the formation of such thermalized  many-body states is
given by the inverse spreading width, $\tau_{\rm th}=
\hbar/\Gamma_{\rm spr}$ \cite{anderson,GM-GW98}. The quantity
$\Gamma_{\rm spr}$ also characterizes the width of the
distribution of the expansion coef\/f\/icients of the many-body
eigenstates over a noninteracting mean-f\/ield basis
\cite{wigner55:57,wigner72,anderson,GM-GW98}.

There is another way to interpret spreading width if there exists
a classical analog to the system. Then it is given by the width of
the energy distribution of trajectories determined
 by the Hamiltonian with interaction, measured in terms of the independent particle Hamiltonian.
  This distribution also provides a description of semiclassical wave functions
  \cite{Thomas1} around which the actual wave functions have Gaussian f\/luctuations \cite{Thomas2}.

The question now arises whether phase relations and/or
correlations between these indivi\-dually ergodic, spatially
extended, many-body states in the superposition may still preserve
a memory of the way the system was excited. This question is of
fundamental importance for the study of relaxation phenomena in
nuclear, atomic, molecular and mesoscopic many-body systems, and
for many-qubit quantum computation. In particular, we recently
proposed \cite{floresKS05} that, if phase relaxation is slower
than energy relaxation, this can extend the time for quantum
computing beyond the so called ``quantum-chaos border''
\cite{geor-shep00,shep01}.

To answer this question from f\/irst principles is presently not
possible due to computational limitations. Indeed, in order to
solve the full quantum many-body problem one would require a
many-qubit quantum computer. Therefore, the only currently
available resort to search for possible manifestations of long
phase relaxation is the experiment, and a careful data analysis.
Nuclear systems are an ideal testing ground to study many-body
systems, since nuclear interactions are so strong that external
perturbations can often be neglected. In particular, the analysis
in~\cite{floresKS05,MarcFK05} of the data displaying forward
peaking in angular distributions of evaporation protons from heavy
nuclei in nucleon-induced reactions indicates that phase
relaxation times can be up to f\/ive orders of magnitude longer
than energy relaxation times.

In this paper we analyze another reaction that indicates the
formation of thermalized non-equilibrated matter. This new form of
matter was introduced by one of us \cite{kun94,kun97}. Again this
is revealed by a strong asymmetry around 90$^\circ $~c.m.\ of the
evaporating proton yield, but now in the Bi($\gamma$,p)
photonuclear reaction \cite{dataBi}. The experiment indicates that
in this case the ef\/fect is even more pronounced. We shall see
that here the phase relaxation is about eight orders of magnitude
slower than thermalization, making the possibility of quantum
computation beyond the quantum chaos border an even more
attractive concept.

The present work is a step toward a more realistic situation
because the entrance channel is given by electro-magnetic
interaction. This is more similar to loading mechanisms in
proposed quantum computers.

The present work does not suggest to use photonuclear reactions as
a practical experimental setup for universal quantum computation.
Indeed, one can not perform a universal set of gates for the
considered photonuclear reaction. However the proposed analysis
does demonstrate that phase relaxation can be much longer than
thermalization which, in turn, illustrates the main idea of
quantum computing far beyond the quantum chaos border.

\section[Experimental manifestation of thermalized non-equilibrated matter
in Bi($\gamma$,p) evaporation process]{Experimental manifestation
of thermalized non-equilibrated\\ matter in
Bi($\boldsymbol{\gamma}$,p) evaporation process}

We analyze the proton yield of the Bi($\gamma$,p) photonuclear
reaction produced by 24-MeV brems\-strahlung. In Fig.~1 we present
an angle-integrated photo-proton spectrum scaled with the outgoing
proton energy ${\varepsilon }$ times the cross section
$\sigma_{\rm inv}(\varepsilon )$ of the inverse process of the
capture of the proton with energy ${\varepsilon }$ by the residual
nucleus. This ${\varepsilon }$-dependent inverse cross section is
determined by the penetrability of the Coulomb and centrifugal
barriers and was taken from~\cite{blattweiss91} for a nuclear
radius of $1.5\times 10^{-13} A^{1/3}$ cm, where $A$ is the
nuclear mass number.

\begin{figure}[t]
\centerline{\begin{minipage}{7.5cm}
\centerline{\includegraphics[width=7.5cm]{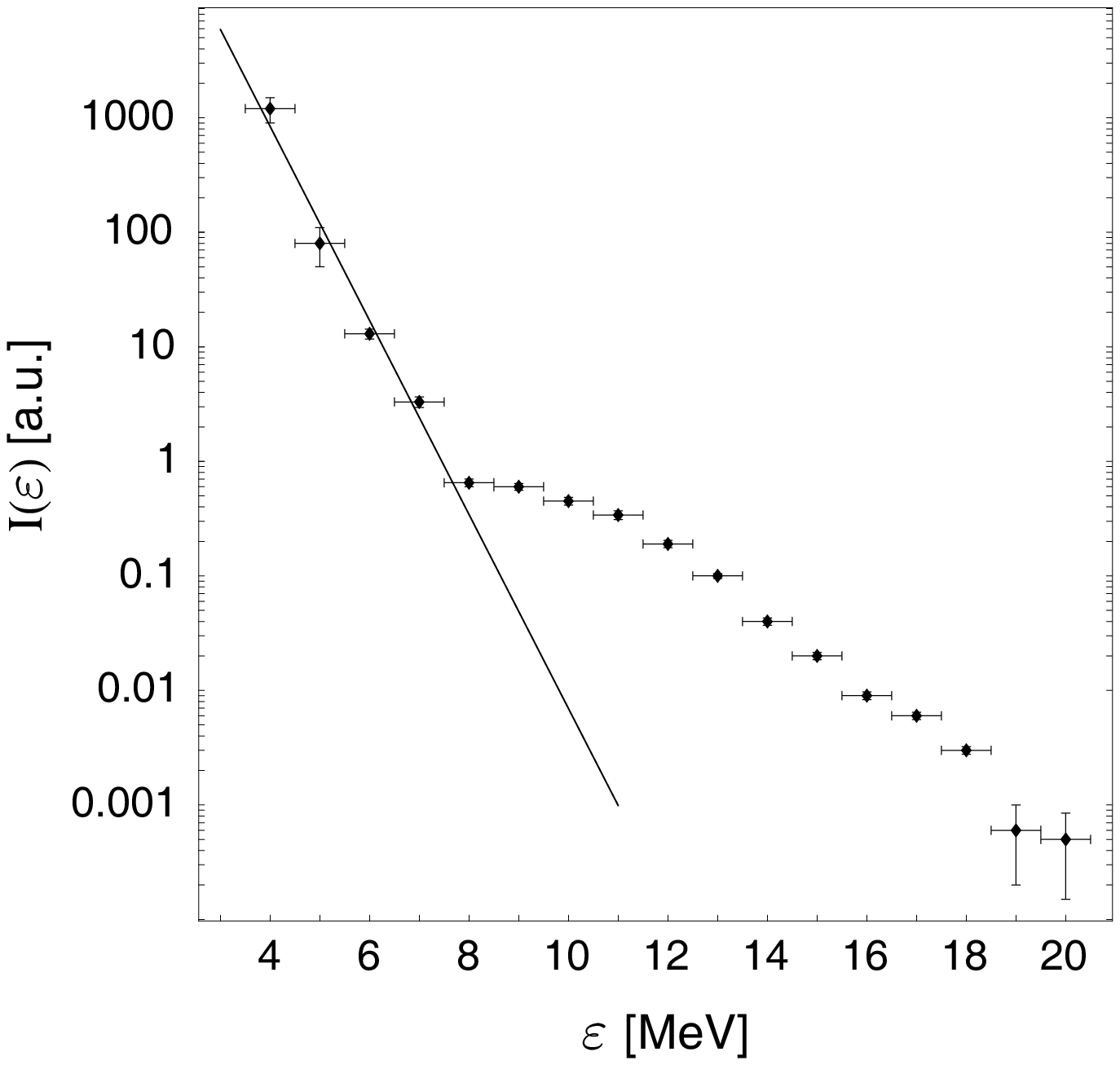}}
\vspace{-2mm} \caption{\label{fig:1} Scaled experimental
angle-integrated spectrum (in arbitrary units) of protons of the
Bi($\gamma$,p) photonuclear reaction produced by 24~MeV
bremsstrahlung~\cite{dataBi}. The line is exponential f\/it of the
scaled spectrum for $\varepsilon \leq 8$ MeV with the slope
(nuclear ``temperature'') of 0.55 MeV (see text).}
\end{minipage}\hfill
\begin{minipage}{7.5cm}
\vspace*{-1mm}
\centerline{\includegraphics[width=7.5cm]{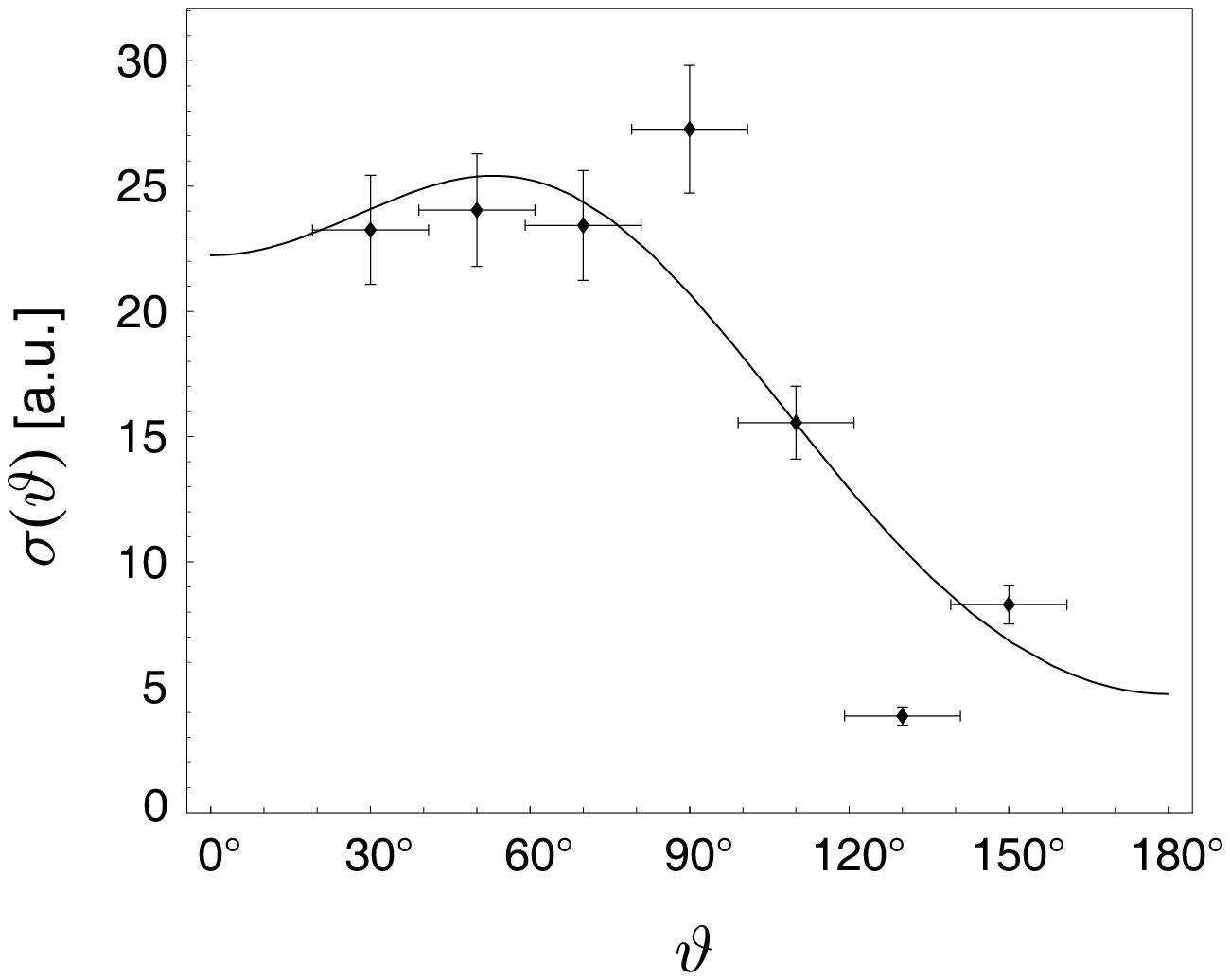}}
\vspace{-3mm} \caption{\label{fig:2} Experimental proton angular
distri\-bu\-tions (in arbitrary units) from the Bi($\gamma$,p)
pho\-to\-nuclear reaction for $\varepsilon =2$--8~MeV \cite{dataBi}.
The cur\-ve is a f\/it to the experimental data (see text).}
\end{minipage}}
\vspace{-3mm}
\end{figure}

One can see that, for $\varepsilon \leq 8$ MeV, the scaled
spectrum has an exponential shape with a~slope of 0.55 MeV. This
is characteristic for the decay of thermalized compound nucleus
with a~``temperature''   $T=0.55$ MeV of the residual nucleus. We
evaluate the average excitation energy of the compound nucleus to
be $\bar E^\ast =14$ MeV, i.e.\ slightly above the center of the
giant resonance peak at 13.5~MeV~\cite{Bigiantres64}. For
$\varepsilon =4$~MeV, the average excitation energy of the
residual nucleus is $\bar E_{\rm res}^\ast =6.3$~MeV. The average
number of excitons (particles plus holes)~\cite{blann75} in the
residual nucleus is $\bar n=(2g\bar E_{\rm res}^\ast )^{1/2}\simeq
14$, where $g=A/13$~MeV$^{-1}$ is the level density of independent
particle states. The standard deviation of $n$ is about $(\bar n
/2 )^{1/2}\simeq 2.6$. Therefore, the energy per exciton in the
residual nucleus, i.e.\ the nuclear ``temperature'', is
$T=0.38$--0.55~MeV. This is consistent with the f\/it in Fig. 1
for   $\varepsilon \leq 8$ MeV with $T=0.55$ MeV. Similarly we
f\/ind that the temperature of the Bi compound nucleus with the
excitation energy 14 MeV is 0.57--0.78 MeV which is roughly
consistent with the f\/it in Fig.~1. This implies
 thermalization of both compound and residual nuclei~\cite{blattweiss91}.

In Fig.~2 we present experimental proton angular distributions
from the Bi($\gamma$,p) photonuclear reaction for $\varepsilon
=2$--8~MeV \cite{dataBi}. One can see that, in spite of a complete
energy relaxation in the thermalized compound nucleus, the angular
distributions are strongly asymmetric about~90$^\circ $, i.e.\
 memory of the direction of the incident $\gamma$-ray beam is clearly retained.

\section{Determination of the cross symmetry phase relaxation width}

In order to identify anomalously slow phase relaxation behind the
forward peaking of the eva\-porating protons we make use of a
standard formula for double dif\/ferential cross section in
$\gamma$-ray induced reactions \cite{simon53}:
\begin{gather}
\sigma (\theta)=(\lambda^2/8\pi )(2I+1)^{-1}\sum {S_{\alpha L_1
p;\alpha^\prime l_1^\prime s^\prime }^{J_1\pi_1}}(E)^\ast
S_{\alpha L_2 p;\alpha^\prime l_2^\prime s^\prime }^{J_2\pi_2}(E) \nonumber\\
\phantom{\sigma (\theta)=}{} (L_1{~}1{~}-1{~}1|L_1{~}1{~}l_1{~}0)
(L_2{~}1{~}-1{~}1|L_2{~}1{~}l_2{~}0)
\Delta(l_1,p)\Delta(l_2,p) \nonumber\\
\phantom{\sigma (\theta)=}{} [(2J_1+1)(2J_2+1)]^{1/2}
i^{L_2-L_1+l_1-l_2} (-1)^{s^\prime-I-L+L_1-L_2+1}
W(J_1L_1J_2L_2;IL) \nonumber\\
\phantom{\sigma (\theta)=}{} Z(l_1L_1l_2L_2;1{~}L) Z(l_1^\prime
J_1l_2^\prime J_2;s^\prime{~}L) P_L(\theta ),\label{eq1}
\end{gather}
where the sum is over $L_1$, $L_2$, $l_1$, $l_2$, $J_1$, $J_2$,
$\pi_1$, $\pi_2$, $l_1^\prime$, $l_2^\prime$, $s^\prime$, $p$ and
$L$. Here $\lambda$ is the wave length of the electromagnetic
radiation, $E$  the total energy of the system, $I$  the spin of
the target nucleus, and $\alpha$, $\alpha^\prime $ are microstates
of the reaction partners in the entrance and exit channels,
accordingly. $J_1$, $J_2$ are the total spins of the compound
nucleus, $\pi_1$, $\pi_2$ its parities, $L_1$, $L_2$  the total
angular momenta of multipoles,  $l_1$, $l_2$ the orbital momenta
in the entrance and $l_1^\prime$, $l_2^\prime$ in the exit
channel, and $s^\prime$ is the channel spin in the exit channel.
The symbol $p$ is def\/ined to have the value 0 for magnetic
radiation and 1 for electric radiation. For $p=0$,
$\Delta(l_{1(2)},p)=\delta_{l_{1(2)},L_{1(2)}}$ and, for $p=1$,
 $\Delta(l_{1(2)},p)=\delta_{l_{1(2)},L_{1(2)}\pm 1}$.
In equation \eqref{eq1}, $(\cdot|\cdot)$ denote the
Clebsch--Gordan coef\/f\/icients, $W$ are the Racah
coef\/f\/icients def\/ined in~\cite{racah42}, and the $Z$
coef\/f\/icients are def\/ined in~\cite{blatt-bied52}. $S$-matrix
elements are in a mixed representation in which the entrance
channel states are in the multipole representation, and the exit
channel states are in the channel spin representation.

In what follows we use a symplif\/ied form of equation
\eqref{eq1}. We neglect the proton spin  in the exit channel so
that the spin of the residual nucleus, $I^\prime $, is equal to
the exit channel spin. We neglect the target spin in the entrance
channel, $I=0$. Therefore, $J_{1(2)}=L_{1(2)}$. We take into
account the proton orbital angular momenta $l_{1,2}^\prime \leq 2$
only since, for $l_{1,2}^\prime\geq 3$, evaporating protons are
signif\/icantly sub-barrier due to the centrifugal and Coulomb
barriers. We consider electric dipole ($L_{1,2}=1$) and quadrupole
($L_{1,2}=2$) radiation only. As a result, equation \eqref{eq1}
takes the form
\begin{gather}
\sigma (\theta)=(\lambda^2/8\pi )\sum {S_{\alpha L_1
;\alpha^\prime l_1^\prime }^{L_1}}(E)^\ast
S_{\alpha L_2 ;\alpha^\prime l_2^\prime }^{L_2}(E) \nonumber\\
\phantom{\sigma (\theta)=}{}(L_1{~}1{~}-1{~}1|L_1{~}1{~}l_1{~}0)
(L_2{~}1{~}-1{~}1|L_2{~}1{~}l_2{~}0)
\delta_{l_1, L_1\pm 1 }\delta_{l_2, L_2\pm 1} \nonumber\\
\phantom{\sigma (\theta)=}{}i^{L_2-L_1+l_1-l_2} (-1)^{I^\prime +1}
Z(l_1L_1l_2L_2;1{~}L) Z(l_1^\prime L_1l_2^\prime L_2;I^\prime{~}L)
P_L(\theta ),\label{eq2}
\end{gather}
where the sum is taken over $L_1$, $L_2$, $l_1$, $l_2$,
$l_1^\prime$, $l_2^\prime$, $I^\prime $ and $L$. In
equation~\eqref{eq2}, the states with dif\/ferent total spins,
$L_1\neq L_2$, correspond to opposite parities. For $L_1=L_2$
$(L_1\neq L_2)$, the sum in equation \eqref{eq2} includes the
terms with even (odd) values of $|l_1^\prime -l_2^\prime |$ only.

The key element in a description of the asymmetry of the
evaporating protons is a correlation between f\/luctuating
$S$-matrix elements with {\it different total spins} $L_1\neq L_2$
($L_1=1$ and $L_2=2$ or $L_1=2$ and $L_2=1$) and {\it parities}
\cite{kun94,kun97}:
\begin{gather}\label{eq3}
\langle{S_{\alpha L_1 ;\alpha^\prime l_1^\prime }^{L_1}}(E)^\ast
S_{\alpha L_2 ;\alpha^\prime l_2^\prime }^{L_2}(E)\rangle =
\frac{\left[\langle|{S_{\alpha L_1 ;\alpha^\prime l_1^\prime
}^{L_1}}(E)|^2\rangle \langle|S_{\alpha L_2 ;\alpha^\prime
l_2^\prime }^{L_2}(E)|^2\rangle\right]^{1/2}}{
(1+\beta/\Gamma_{\rm cn})},
\end{gather}
where the brackets $\langle \cdot \rangle$ stand for ensemble
averaging which is equivalent to the energy ($E$) ave\-raging
under the ergodicity condition. The physical meaning of the spin
and parity of\/f-diagonal (``cross symmetry'') phase relaxation
width $\beta$, introduced in~\cite{kunPLB93,kun94,kun97}, is the
characteristic inverse time over which the interference between
the states with dif\/ferent total spins and parities does not
vanish upon the energy averaging and therefore the memory of the
direction of the initial beam is preserved~\cite{kun94,kun97}. If
this phase memory time $\tau_{ph}=\hbar/\beta$ is about as long or
longer than the average life time $\hbar/\Gamma_{\rm cn}$ of the
compound nucleus then the evaporating protons are emitted
asymmetrically about 90$^\circ$ c.m., i.e.\ the memory about the
direction of the initial beam is retained. However, if the phase
memory time is much shorter than the average life time of the
compound nucleus then the spin and parity of\/f-diagonal
correlations vanish, memory on the direction of the initial beam
is lost leading to the conventional Bohr picture of compound
nucleus with the symmetric about  90$^\circ$ c.m.
 angular distributions.
 Therefore, deviation
of the angular distributions of the evaporating particles from the
symmetry around 90$^\circ$ c.m.\ indicates that $\beta$ is smaller
or comparable to $\Gamma_{\rm cn}$.

For the orbital momentum of\/f-diagonal correlation between
$S$-matrix elements with the same total spins and parities we have
\cite{kun94,kun97}
\begin{gather}
\langle{S_{\alpha L ;\alpha^\prime l_1^\prime  }^{L}}(E)^\ast
S_{\alpha L ;\alpha^\prime l_2^\prime }^{L}(E)\rangle=
[\langle|{S_{\alpha L ;\alpha^\prime l_1^\prime
}^{L}}(E)|^2\rangle \langle|S_{\alpha L ;\alpha^\prime l_2^\prime
}^{L}(E)|^2\rangle]^{1/2}.\label{eq4}
\end{gather}
Both the equations \eqref{eq3} and \eqref{eq4} ref\/lect a strong
correlation between the partial width amplitudes and, as a result,
between $S$-matrix elements with dif\/ferent orbital momenta
($l_1^\prime\neq l_2^\prime $) of evaporating protons referred
to~\cite{kun94,kun97} as the continuum correlation. Note that such
a strong correlation between reduced width amplitudes
corresponding to the same total spin and parity values but
dif\/ferent orbital momenta was experimentally revealed for a
number of compound nuclei in the regime of isolated
resonances~\cite{mitchell85}.

The statistical model \cite{blattweiss91} yields
\begin{gather}
\langle|{S_{\alpha L ;\alpha^\prime l^\prime }^{L}}(E)|^2\rangle
=T_\alpha^L T_{\alpha^\prime l^\prime }^L/\sum_{\alpha^\prime
l^\prime } T_{\alpha^\prime l^\prime }^L.\label{eq5}
\end{gather}
Here, $T_\alpha^L\equiv T^L$ are the entrance channel transmission
coef\/f\/icients for the formation of the compound nucleus with
the total spins $L=1$ and $L=2$ due to the absorption of electric
dipole and quadrupole radiation, accordingly. The exit channel
transmission coef\/f\/icients are assumed to be independent of the
compound nucleus spin $L$ and the spin of the residual nucleus
$I^\prime $ \cite{blattweiss91}, $T_{\alpha^\prime l^\prime
}^L\equiv T^{l^\prime}$.

We use equations \eqref{eq2}, \eqref{eq3}, \eqref{eq4} and
\eqref{eq5} for the analysis the experimental angular
distributions in Fig.~2. We f\/ind that, upon energy averaging and
summation over microstates of the residual nucleus, apart from the
overall normalization constant, a shape of the angular
distributions depends on the four parameters: $A=T^{L=2}/T^{L=1}$,
$B=T^{l^\prime =1}/T^{l^\prime =0}$,
 $C=T^{l^\prime =2}/T^{l^\prime =0}$, and~$\beta/\Gamma_{\rm cn}$.

From the f\/it of the experimental angular distributions in Fig.~2 we obtain:
$A=0.082$, $B=0.47$, $C=0.37$ and $\beta/\Gamma_{\rm cn}=0.11$.
The compound nucleus' decay width $\Gamma_{\rm cn}$ for Bi with an
excitation energy of 14~MeV can be estimated from the systematics
in Fig.~7 of~\cite{eric-m-k66}, which provides a good description
of the experimentally determined $\Gamma_{\rm cn}$ for a wide
range of mass numbers. From this estimation we obtain $\Gamma_{\rm
cn}\simeq 0.1$~eV what yields in turn $\beta\simeq 0.01$~eV. At
the same time, the standard nuclear physics estimate for the
spreading width of Bi nucleus with the excitation energy 14 MeV is
about 2 MeV (see Fig.~2.1 in~\cite{AWM75}). This is close to
another estimate of $\Gamma_{\rm spr}$ as a width of a dipole
giant resonance \cite{anderson}, which is about 4.5 MeV for
Bi~\cite{Bigiantres64}. It corresponds to an exponentially large
ef\/fective dimension of Hilbert space given by $N_{\rm eff}\simeq
\Gamma_{\rm spr}/D\simeq 10^{16}$, where $D\simeq 10^{-16}$ MeV is
the average level spacing of the Bi compound nucleus with the
excitation energy 14 MeV. Note that the total spin and parity
of\/f-diagonal $S$-mat\-rix correlations for evaporation processes
were justif\/ied in~\cite{kun97} in the limit $N_{\rm
eff}\to\infty $. For the cross symmetry phase relaxation time much
longer than the energy relaxation time, the formalism also leads
to (i) quantum-classical transition \cite{KVG01,BenetKW05}, stable
coherent rotation \cite{KunVagVor99,KunRobVag99,KunChVG02},
Schr\"odinger cat states \cite{KunBChGrH03,BKWD05} in complex
collisions in the regime of strongly overlapping resonances of the
intermediate system, and (ii) spontaneous correlations,
non-equilibrium phase transitions~\cite{KunAIP01} and anomalous
sensitivity in f\/inite highly excited many-body
systems~\cite{KunPRL00}.

It should be noted that while we have been able to determine an
anomalously small value of $\beta$ from the data analysis, its
theoretical evaluation is currently an open problem. Therefore
more theoretical insight is needed for a deeper understanding of
the ef\/fect of anomalously slow cross symmetry phase relaxation
in highly excited quantum many-body systems.

\section{An illustration of the idea of quantum computing far\\ beyond the quantum chaos boarder}

The idea of quantum computing on a time scale  $\tau_{\rm
ph}=\hbar/\beta\simeq 6\times 10^{-14}$ sec., which is much longer
than the thermalization time $\hbar /\Gamma_{\rm spr}\simeq
3\times 10^{-22}$ sec., can be illustrated as
follows~\cite{floresKS05}. Consider the Bi($\gamma$,p) compound
nucleus reaction to be the quantum protocol. The single-particle
basis is the quantum register. The entrance channel represents the
loading process. The measurement of the angular distribution of
the yield of evaporating protons plays the role of the readout. In
accordance with the standard criterion \cite{geor-shep00,shep01},
a quantum computer melts down at $t\geq \hbar /\Gamma_{\rm spr}$.
If so, then any information about specif\/ic features of the
loaded state (entrance channel) is lost while the protocol is
executed. Yet, as was shown in the previous section, our ``read
out'' (the angular distribution of evaporation protons) shows that
our ``protocol'' yields non-trivial results eight orders of
magnitude later than the time scale $\hbar/\Gamma_{\rm spr}\simeq
3\times 10^{-22}$ sec. for the onset of quantum chaos in our
``quantum computer''. These results depend on the ``loading'',
i.e.\ the direction of the incident beam.

In order to demonstrate that the obtained phase memory time
$\tau_{\rm ph}=\hbar/\beta\simeq 6\times 10^{-14}$ sec.\ is indeed
a very long time scale of the system we note that during this time
a nucleon at the Fermi energy crosses the Bi compound nucleus
about 10$^8$ times. Yet, this phase memory time is still about
eight orders of magnitude
 shorter than the Heisenberg time $\hbar/D\simeq 10^{-5}$
sec., where $D\simeq 10^{-16}$ MeV is the average level spacing of
the Bi compound nucleus at an excitation energy of 14 MeV.

Clearly, the spectrum of the Bi compound nucleus in the above
considered reaction is not resolved since $\Gamma_{\rm cn}/D\simeq
10^9$, i.e.\ resonances of the compound nucleus are strongly
overlapping. Therefore, all exponentially large information hidden
in the Hilbert space is not available. Ho\-we\-ver we have
demonstrated that useful information, such as the phase relaxation
time, still can be obtained even though the quantum protocol time
is about nine orders of magnitude shorter than the Heisenberg time
for our ``quantum computer''. This is in accord with a similar
observation in~\cite{shep01}.

\section{Conclusions}

We have demonstrated that thermalization in highly excited quantum
many-body system does not necessarily mean a complete loss of
memory
 of the way the system was formed.
Manifestation of such a thermalized non-equilibrated matter has
been revealed from the analysis of a strong asymmetry around
90$^\circ $ c.m.\ of the evaporating proton yield in the
Bi($\gamma$,p) photonuclear reaction. We have shown that
thermalized non-equilibrated matter can exist for time spans of
$\sim 6\times 10^{-14}$~sec., which is eight orders of magnitude
longer than thermalization time in this example. This indicates
that long lived transient states can exist in many-body systems
with exponentially large dimensions of Hilbert space. If a quantum
computer  with a large number of qubits $n\simeq 100$--1000 can be
brought into such a state, this may provide a solution for the
scaling problem which is one of the central challenges of quantum
information~\cite{Nielsen}.

\subsection*{Acknowledgments}

S. Kun is grateful to the organizers of the 6th International
Conference
 ``Symmetry in Nonlinear Mathematical Physics'',
 which took place in his beautiful home city Kyiv in June 2005,
 for the kind invitation and the opportunity to give a talk.
We acknowledge f\/inancial support from PAPIIT project 10803 and
CONACyT project 41000-F. M.~Bienert is supported by a Feodor-Lynen
 fellowship of the Alexander-von-Humboldt foundation.

\LastPageEnding


\begin{thebibliography}{99}

\footnotesize

\bibitem{wigner55:57}
Wigner E.P., Characteristic vectors of bordered matrices with
inf\/inite dimensions, {\it Ann. Math.}, 1955, V.62, N~3,
548--564.\\
Wigner E.P., Characteristic vectors of bordered matrices with
inf\/inite dimensions II, {\it Ann. Math.}, 1957, V.65, N~2,
203--207.

\bibitem{wigner72}
Wigner E., Statistical properties of nuclei, Editor J.B.~Garg, New
York~-- London, Plenum Press, 1972, p.~11.

\bibitem{anderson}
Anderson P.W., Basic notions of condensed matter physics, {\it
Frontiers in Physics}, Vol.~55, The Benjamin-Cummings, 1984,
71--72.

\bibitem{GM-GW98}
Guhr T., M\"uller-Groeling A., Weidenm{\" u}ller H.A., Random
matrix theories in quantum physics: common concepts, {\it Phys.
Rep.}, 1998, V.299, N~4--6, 189--425 (and references therein); cond-mat/9707301.

\bibitem{Thomas1}
Benet L., Izrailev F.M., Seligman T.H., Su{\'a}rez-Moreno A.,
Semiclassical properties of eigenfunctions and occupation number
distribution for a model of two interacting particles, {\it Phys.
Lett.~A}, 2000, V.277, N~2, 87--93; chao-dyn/9912035.

\bibitem{Thomas2}
Benet L., Flores J., Hernandez-Saldana H., Izrailev F.M., Leyvraz
F., Seligman T.H., Fluctuations of wavefunctions about their
classical average {\it J. Phys. A: Math. Gen.}, 2003, V.36, N~5,
1289--1297; nlin.CD/0207039.

\bibitem{floresKS05}
Flores J., Kun S.Yu., Seligman T.H., Slow phase relaxation as a
route to quantum computing beyond the quantum chaos border, {\it
Phys. Rev. E}, 2005, V.72, N~1, 017201, 4 pages.

\bibitem{geor-shep00}
Georgeot B., Shepelyansky D.L., Quantum chaos border for quantum
computing, {\it Phys. Rev. E}, 2000, V.62, N~3, 3504--3507; quant-ph/9909074.

\bibitem{shep01}
Shepelyansky D.L., Quantum chaos and quantum computers, in
Proceedings of Nobel Symposium on Quantum Chaos 2000, {\it Phys.
Scripta}, 2001, V.90, 112--120; quant-ph/0006073.

\bibitem{MarcFK05}
Bienert M., Flores J., Kun S.Yu., Experimental proposal for
accurate determination of the phase relaxation time in highly
excited quantum many-body systems, nucl-ex/0508020.

\bibitem{kun94}
Kun S.Yu., Novel approach to angular distributions in precompound
reactions: Does the Bohr hypothesis always work?, {\it Z. Phys.
A}, 1994, V.348, N~2, 273--279.

\bibitem{kun97}
Kun S.Yu., Statistical reactions with memory and
thermalized-nonequilibrated nuclear states,
 {\it Z. Phys. A}, 1997, V.357, N~2, 255--269.

\bibitem{dataBi}
Toms M.E., Stephens W.E., Photoprotons from In, Ce, and Bi, {\it
Phys. Rev.}, 1953, V.92, N~2, 362--366.

\bibitem{blattweiss91}
Blatt J.M., Weisskopf V.F., Theoretical nuclear physics, New York,
Dover Publications, Inc., 1991.

\bibitem{Bigiantres64}
 Harvey R.R., Caldwell J.T., Bramblett R.L., Fultz S.C., Photoneutron cross sections of Pb$^{206}$,
 Pb$^{207}$, Pb$^{208}$, and Bi$^{209}$,
{\it Phys. Rev.}, 1964, V.136, N~1, B126--B131.

\bibitem{blann75}
Blann M., Preequilibrium decay, {\it Annu. Rev. Nucl. Sci.}, 1975,
V.25, 123--166.

\bibitem{simon53}
Simon A., Theory of polarized particles and gamma rays in nuclear
reactions, {\it Phys. Rev.}, 1953, V.92, N~4, 1050--1060.

\bibitem{racah42}
Racah G., Theory of complex spectra. I, {\it Phys. Rev.}, 1942,
V.61, N~3--4, 186--197.\\
Racah G., Theory of complex spectra.~II, {\it Phys.
Rev.}, 1942, V.62, N 9--10, 438--462.

\bibitem{blatt-bied52}
Blatt J.M., Biedenharn L.C., The angular distributions of
scattering and reaction cross sections, {\it Rev. Mod. Phys.},
1952, V.24, N~4, 258--272.

\bibitem{kunPLB93}
Kun S.Yu., Ericson f\/luctuations for the f\/inite spin relaxation
times,
 {\it Phys. Lett.~B}, 1993, V.319, N~1, 16--22.

\bibitem{mitchell85}
Mitchell G.E., Bilpuch E.G., Shriner J.F., Lane A.M., Amplitude
correlations in nuclear resonance spectroscopy, {\it Phys. Rep.},
1985, V.117, N~1, 1--74.

\bibitem{eric-m-k66}
Ericson T., Mayer-Kuckuk, Fluctuations in nuclear reactions, {\it
Annu. Rev. Nucl. Sci.}, 1966, V.16, 183--206.

\bibitem{AWM75}
Agassi D., Weidenm\"uller H.A., Mantzouranis G., The statistical
theory of nuclear reactions for strongly overlapping resonances as
a theory of transport phenomena,
 {\it Phys. Rep.}, 1975, V.22, N~3, 145--179.

\bibitem{KVG01}
Kun S.Yu., Vagov A.V., Greiner W., Quantum-classical
correspondence in microscopic and mesoscopic complex collisions,
{\it Phys. Rev.~C}, 2001, V.63, N~1, 014608, 5 pages.

\bibitem{BenetKW05}
Benet L., Kun S.Yu., Wang Qi, Ef\/fect of phase relaxation on
quantum superpositions in complex collisions, quant-ph/0503046.

\bibitem{KunVagVor99}
Kun S.Yu., Vagov A.V., Vorov O.K., Coherently rotating
hyperdeformed quasimolecules in $^{12}$C+$^{24}$Mg  scattering?,
{\it Phys. Rev.~C}, 1999, V.59, N~2, R585--R588.

\bibitem{KunRobVag99}
Kun S.Yu., Robson B.A., Vagov A.V., Oscillating-correlated
nonstatistical structures, slow spin decoherence, and
hyperdeformed coherent rotational states in $^{24}$Mg+$^{24}$Mg
and $^{28}$Si+$^{28}$Si scattering, {\it Phys. Rev. Lett.}, 1999,
V.83, N~3, 504--507.

\bibitem{KunChVG02}
Kun S.Yu., Chadderton L.T., Vagov A.V., Greiner W., A new probe
for coherent many-body dynamics: Non\-ergo\-dic molecules in
continuum, {\it Int. J. Mod. Phys.~E}, 2002, V.11, N~4, 273--280.

\bibitem{KunBChGrH03}
Kun S.Yu., Benet L., Chadderton L.T., Greiner W., Haas F.,
Macroscopic quantum superpositions in highly excited strongly
interacting many-body systems, {\it Phys. Rev.~C}, 2003, V.67,
N~1, 011604, 4 pages; quant-ph/0205036.

\bibitem{BKWD05}
Benet L., Kun S.Yu., Wang Qi, Denisov V., Ef\/fect of a
f\/inite-time resolution on Schr\"odinger cat states in complex
collisions, {\it Phys. Lett.~B}, 2005, V.605, N~1--2, 101--105; nucl-th/0407029.

\bibitem{KunAIP01}
Kun S.Yu., Spontaneous coherence and non-equilibrium correlation
phase transitions in microscopic and mesoscopic systems, in
Proceedings of International Conference ``Nonequilibrium and
Nonlinear Dynamics in Nuclear and Other Finite Systems'' (May
21--25, 2001, Beijing), Editors Zhuxia Li, Ke Wu, Xizhen Wu,
Enguang Zhao and F.~Sakata, {\it AIP Conference Proceedings},
Vol.~597, New York, Melville, 2001, 319--326.

\bibitem{KunPRL00}
Kun S.Yu., Sensitivity of nucleus-nucleus cross sections and
atomic-electron ef\/fects in dissipative heavy-ion collisions,
{\it Phys. Rev. Lett.}, 2000, V.84, N~3, 423--426.

\bibitem{Nielsen}
Nielsen M.A., Chuang I.L., Quantum computation and quantum
information, Oxford, Cambridge University Press, 2000.

\end{thebibliography}
\end{document}